LAWRENCE LIVERMORE NATIONAL LABORATORY

LLNL-PROC-564720

# Using the X-FEL to understand X-ray Thomson scattering for partially ionized plasmas

J. Nilsen, W. R. Johnson, K. T. Cheng

July 17, 2012

**Disclaimer**

# Using the X-FEL to understand X-ray Thomson scattering for partially ionized plasmas

Joseph Nilsen[1], Walter R. Johnson[2], and K. T. Cheng[1]

[1]Lawrence Livermore National Laboratory, Livermore, CA 94551
[2]University of Notre Dame, Notre Dame, IN 46556

**Abstract.** For the last decade numerous researchers have been trying to develop experimental techniques to use X-ray Thomson scattering as a method to measure the temperature, electron density, and ionization state of high energy density plasmas such as those used in inertial confinement fusion. X-ray laser sources have always been of interest because of the need to have a bright monochromatic X-ray source to overcome plasma emission and eliminate other lines in the background that complicate the analysis. With the advent of the X-ray free electron laser (X-FEL) at the SLAC Linac Coherent Light Source (LCLS) we now have such a source available in the keV regime. Other X-FEL sources are being built in Germany and around the world. One challenge with X-ray Thomson scattering experiments is understanding how to model the scattering for partially ionized plasmas. Most Thomson scattering codes used to model experimental data greatly simplify or neglect the contributions of the bound electrons to the scattered intensity. In this work we take the existing models of Thomson scattering that include elastic ion-ion scattering and the electron-electron plasmon scattering and add the contribution of the bound electrons in the partially ionized plasmas. Except for hydrogen plasmas almost every plasma that is studied today has bound electrons and it is important to understand their contribution to the Thomson scattering, especially as new X-ray sources such as the X-FEL will allow us to study much higher Z plasmas. Currently most experiments have looked at hydrogen or beryllium. We will first look at the bound electron contributions to beryllium by analysing existing experimental data. We then consider several higher Z materials such as Cr and predict the existence of additional peaks in the scattering spectrum that requires new computational tools to understand. For a Sn plasma we show that the bound contributions changes the shape of the scattered spectrum in a way that would change the plasma temperature and density inferred by the experiment.

## 1 Introduction

Thomson scattering of X-rays is being developed as an important diagnostic technique to measure temperatures, densities, and ionization balance in warm dense plasmas. Glenzer and Redmer [1] have reviewed the underlying theory of Thomson scattering being used in experiments.

In this work we start with the theoretical model proposed by Gregori et al. [2] but with one important difference. We evaluate the Thomson-scattering dynamic structure function using parameters taken from our own average-atom code [3,4]. The average-atom model is a quantum mechanical version of the temperature-dependent Thomas-Fermi model of plasma developed years ago by Feynman et al. and is described in Ref. [5]. It consists of a single ion of charge Z with a total of Z free and bound electrons in a Wigner-Seitz cell that is embedded in a uniform "jellium sea" of free electrons whose charge is balanced by a uniform positive background. This model enables us to consider the contributions from the bound electrons in a self-consistent way for any ion. Other approaches such as Gregori's includes hydrognic wavefunctions with screening factors to approximate the contribution from the bound electrons for a limited number of materials.

In this paper we look first look at the bound electron contributions to beryllium by analyzing existing experimental data. We then consider several higher Z materials such as Cr and predict the existence of additional peaks in the scattering spectrum that requires new computational tools to understand. For a Sn plasma we show that the contributions from the bound electrons can change the shape of the scattered spectrum in a way that would change the plasma temperature and density inferred by the experiment.

## 2 Validating the average-atom code for Be experiments

Numerous experiments have been done to look at Thomson scattering for low-Z materials such as hydrogen and beryllium. We examine one particular experiment[6] done at the Omega laser facility that looked at Thomson scattering in the forward scattering direction (40 degrees) off solid Be with a Cl Ly-$\alpha$ source at 2963 eV. Figure 1 shows the measured spectrum as the noisy dotted line. An electron temperature of 18 eV, ion temperature of 2.1 eV, and density of 1.647 g/cc is used in the average-atom model to give an electron density of 1.8 $\times 10^{23}$ per cc, in agreement with the analysis in Ref. [6]. The dashed line shows the experimental source function from the Cl Ly-$\boldsymbol{\alpha}$ line. Because of satellite structure we approximate the source by 3 lines: a Cl Ly-$\boldsymbol{\alpha}$ line at 2963 eV with amplitude 1 and two satellites at 2934 and 2946 eV with relative amplitudes of 0.075 and 0.037 respectively. Using the 3 weighted lines to do the Thomson scattering calculation, we calculate the scattering amplitude for Thomson scattering (solid line) and compare against the experimental data (dotted line). We observe excellent agreement within the experimental noise. Contributions from the bound 1**s** electrons have a threshold at 2876 eV that are beyond the range of the data shown in the figure.

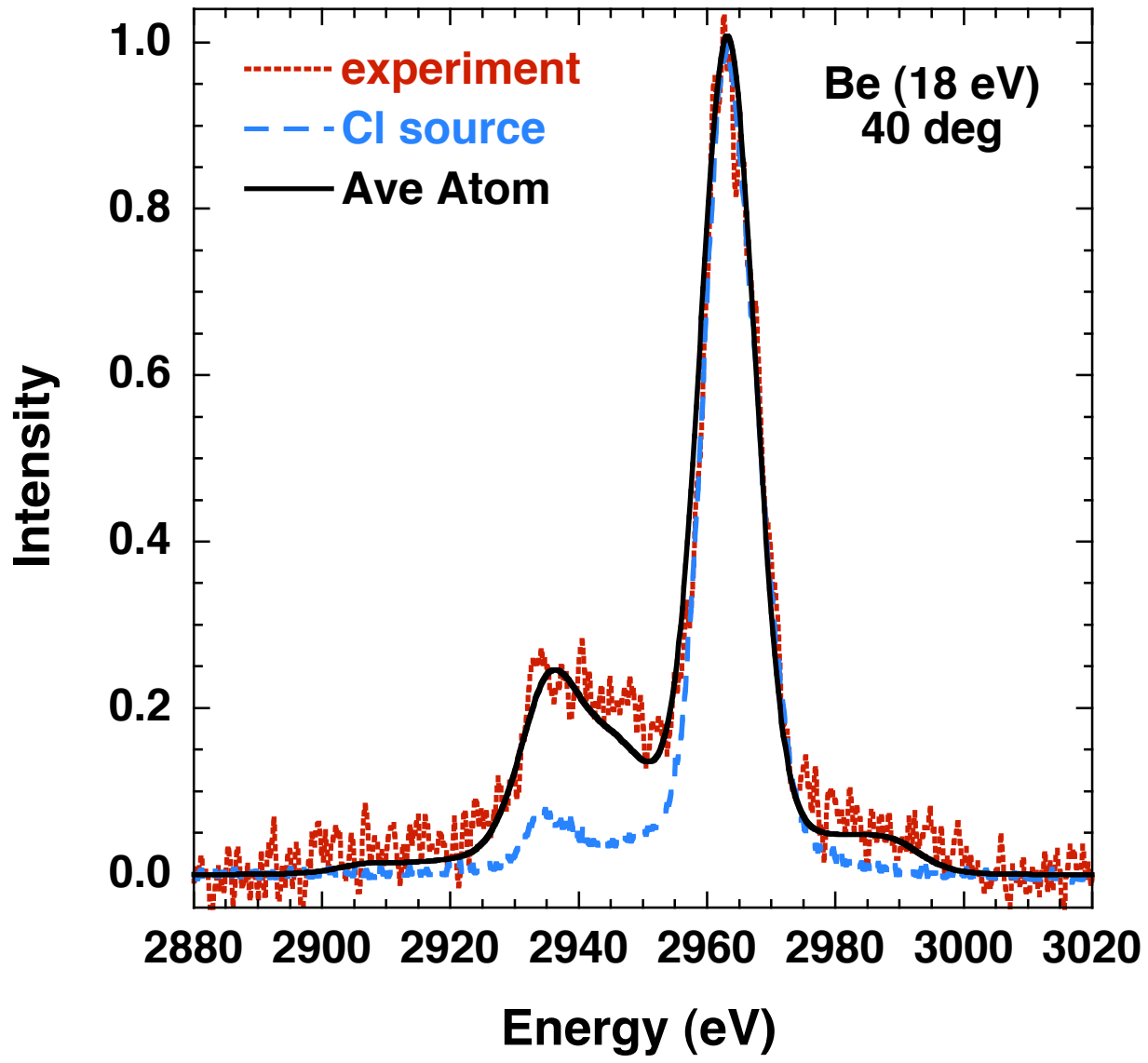

**Fig. 1.** Intensity vs photon energy for scattering of a Cl Ly-α X-ray off Be at 40°. Dotted line is experimental data (see Ref. 6) while solid line is calculation for an electron temperature of 18 eV. The dashed line shows the Cl X-ray source.

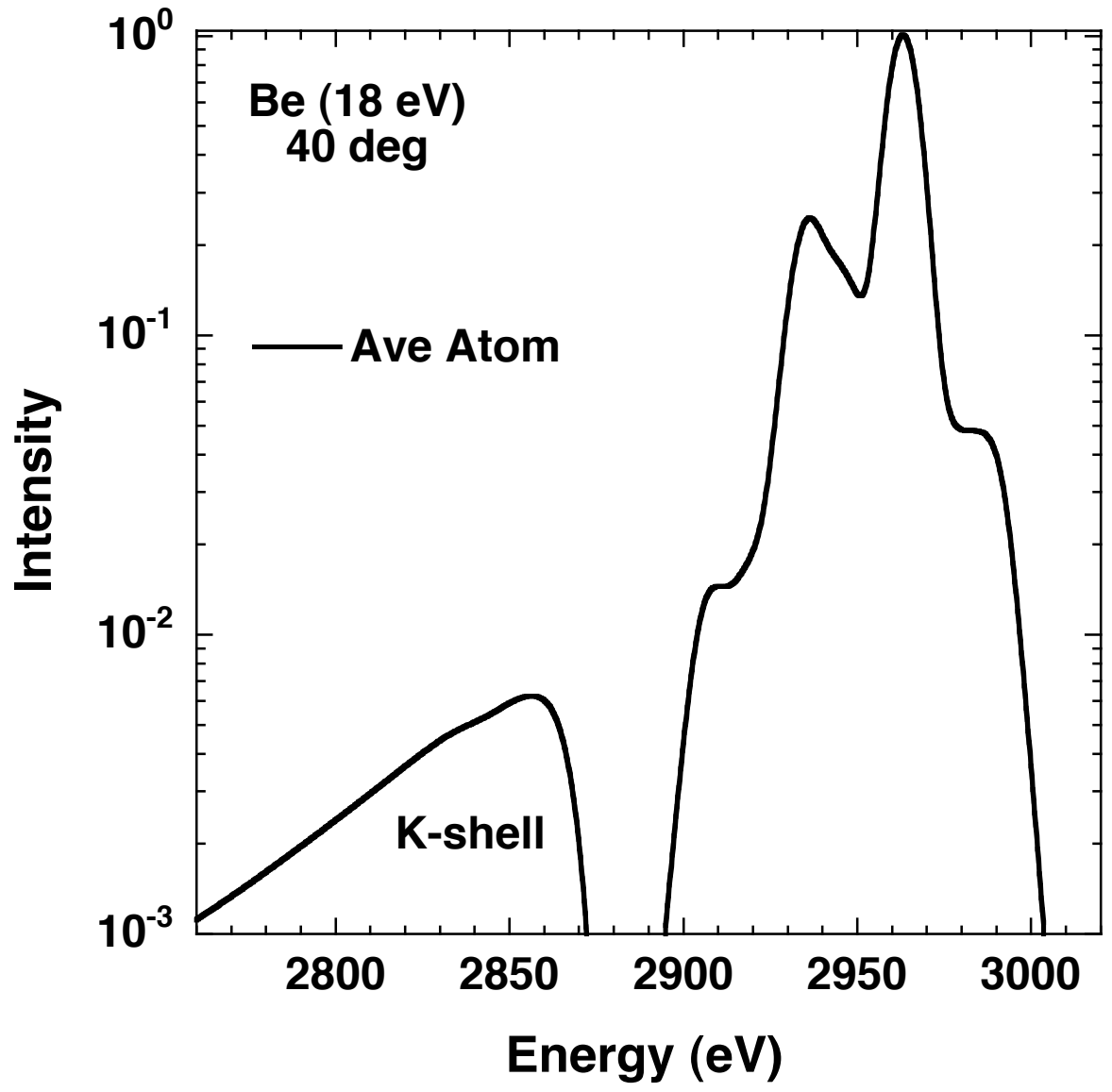

**Fig. 2.** Intensity vs photon energy for calculation of scattering of a Cl Ly-α X-ray off Be at 40° for an electron temperature of 18 eV. The contribution from the K-shell bound electrons is shown.

For Be under the conditions just described the L-shell is completely stripped but the K-shell is 97% occupied. The average-atom code has $Z_f = 1.647$ for the number of free electrons in the jellium sea outside the Wigner-Seitz cell. The binding energy of the K-shell electron is 86.8 eV which means the threshold for seeing the bound state contribution should begin at 2963 – 86.8 = 2876.2 eV. To understand the contributions of the bound K-shell electrons to the Thomson scattering we expand the energy scale in Fig. 1 and re-plot it on a log-scale looking just at the average-atom calculation. In Fig. 2 we observe a K-shell contribution that is about 40 times weaker than the low-energy plasmon peak so it would be very difficult to observe in a laser-plasma experiment given the experimental noise.

## 3 Modeling Cr and Sn experiments

If we now consider higher-Z materials such as Cr and Sn at an electron temperature of 10 eV the bound state contribution can be very important. Following Refs. [4,6] an ion/electron temperature ratio of 0.1 is assumed to reduce the central elastic scattering peak. The average-atom code predicts that solid density Cr (7.19 g/cc) heated to 10 eV has 6.2 continuum electrons in the Wigner-Seitz cell. This makes it a closed Ar-like core with nearly fully occupied 3s and 3p subshells with binding energies of 57.7 and 30.7 eV, respectively. The code predicts $Z_f = 2.92$ and an electron density of $2.4 \times 10^{23}$ per cc. Figure 3 shows the scattered intensity versus photon energy for calculations done with (solid line) and without (dashed line) the contribution of the bound electrons for a 4750 eV X-ray source scattered at 40° off the Cr. Without the bound electrons we predict the central elastic scattering peak and the plasmon peaks. When we include the effect of the bound electrons we predict a very strong scattering peak that is downshifted about 40 eV from the central elastic peak due to the 3p electrons and a weaker 3s peak.

For solid density Sn (7.3 g/cc) at 10 eV we look at the case of the 2960 eV X-ray source scattering in the backward direction at 130°. The average-atom code predicts that warm dense Sn has 4.4 continuum electrons with an average occupation of 8.8 for the 4d electrons and 0.8 for the 5s electrons outside a closed Kr-like core with all the lower orbital subshells fully occupied. We predict $Z_f = 3.37$ which gives an electron density of $1.25 \times 10^{23}$ per cc. The binding energies of the 4d and 5s electrons are 22.4 and 2.14 eV, respectively, which means the effect of the 5s electron will be hidden under the elastic scattering peak. Fig. 4 plots the scattering intensity versus photon energy for calculations done with (solid line) and without (dashed line) the contribution of the bound electrons. For backward scattering the scattering is no longer in the collective regime and the distinct plasmon peaks are replaced by a broader scattering structure. The contribution from the bound 4d electrons is now an additional large broad feature that lies on top of the Thomson scattering and would make it easy to misinterpret the spectrum if one did not understand the contribution of the bound electrons.

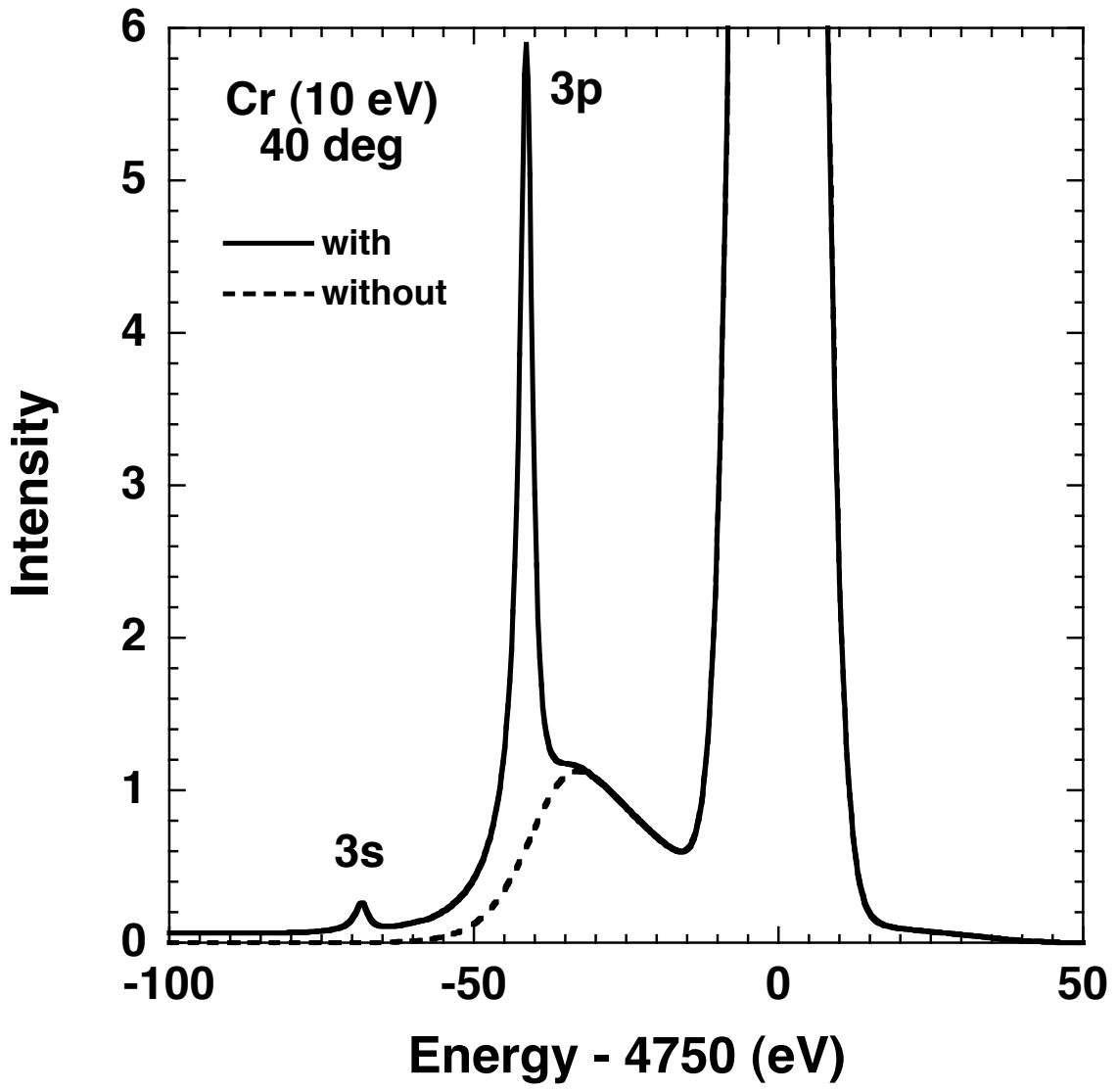

**Fig. 3.** Intensity vs photon energy for calculation of scattering of a 4750 eV X-ray line off 10 eV Cr at 40°. The case where the 3p and 3s bound electron contributions are included is shown by solid line.

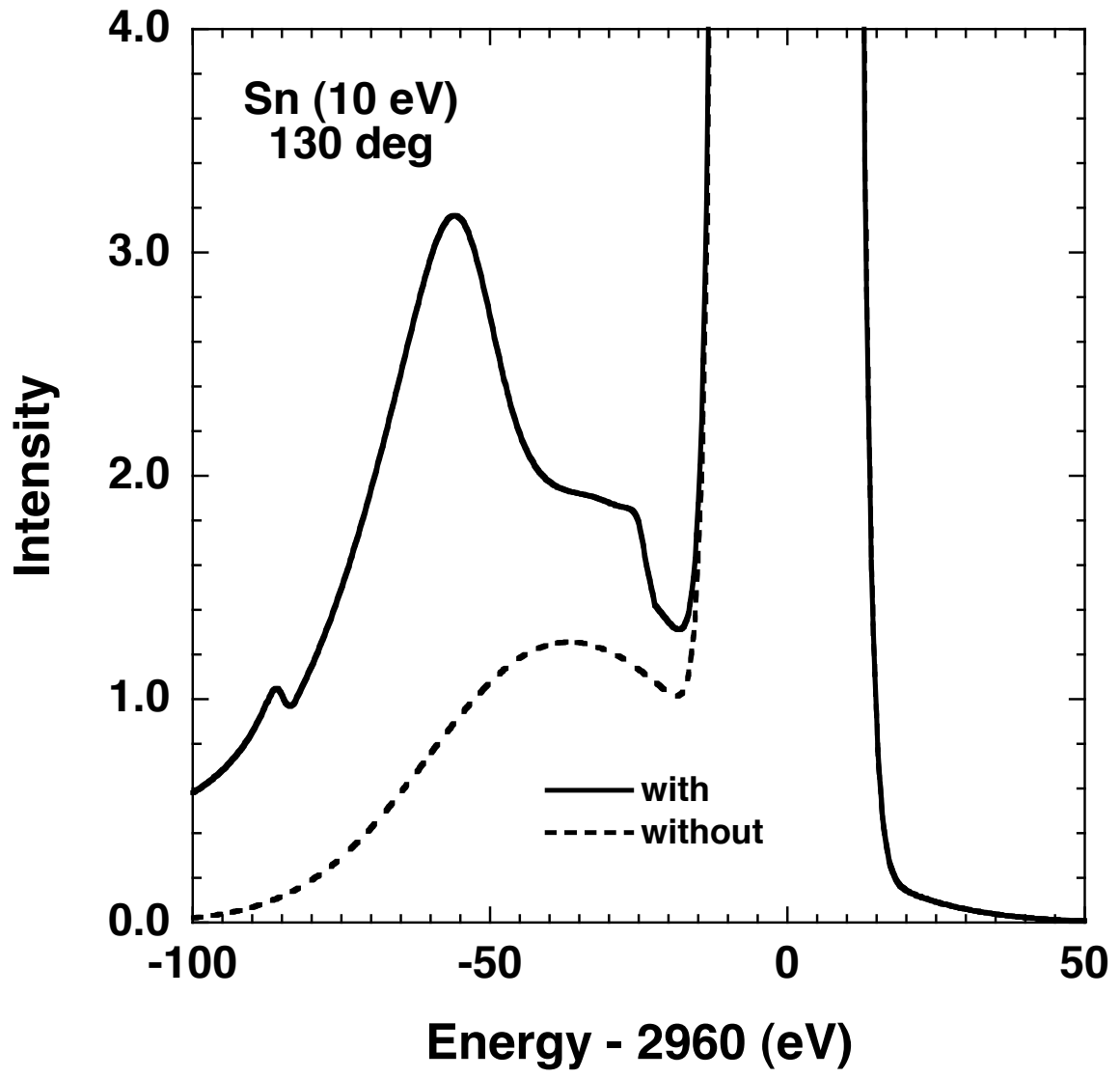

**Fig. 4.** Intensity vs photon energy for calculation of scattering of a 2960 eV X-ray line off Sn at 130° for an electron temperature of 10 eV. The strong contribution from the 4d bound electrons are apparent for the case where the bound electrons are included (solid line) when compared with case without bound electrons (dashed line).

## 4 Conclusions

The availability of bright monochromatic X-ray line sources from X-ray free electron laser facilities opens up many new possibilities to use Thomson scattering as an important diagnostics technique to measure the temperatures, densities, and ionization balance in warm dense plasmas.

Current attempts to model Thomson scattering tend to use very simplified models, if anything, to model the effect of the bound electrons on the measured scattered intensity. Our approach here is to evaluate the Thomson-scattering dynamic structure function using parameters taken from our own average-atom code [3,4]. This model enables us to consider contributions from the bound electrons in a self-consistent way for any ion.

In this paper we validate our average-atom based Thomson scattering code by comparing our model against existing experimental data for Be near solid density and temperature near 18 eV. For solid density Cr at 10 eV we predict the existence of additional peaks in the scattering spectrum that requires new computational tools to understand. We also analyse solid density Sn at 10 eV and show that the contributions from the bound electrons can change the shape of the scattered spectrum in a way that would change the plasma temperature and density inferred by the experiment.

**Acknowledgements.** This work was performed under the auspices of the U.S. Department of Energy by Lawrence Livermore National Laboratory under Contract DE-AC52-07NA27344.